\documentclass[12pt,preprint]{aastex}
\shorttitle{Orbital Inclination of Iapetus}
\shortauthors{Nesvorn\'y et al.}
\begin{document}
\title{Excitation of the Orbital Inclination of Iapetus\\ during Planetary Encounters}
\author{David Nesvorn\'y$^1$, David Vokrouhlick\'y$^{1,2}$, Rogerio Deienno$^{1,3}$
\and Kevin J. Walsh$^1$}
\affil{(1) Department of Space Studies, Southwest Research Institute, 1050 Walnut St., \\Suite 300, 
Boulder, CO 80302, USA} 
\affil{(2) Institute of Astronomy, Charles University, V Hole\v{s}ovi\v{c}k\'ach 2, \\
180 00 Prague 8, Czech Republic}
\affil{(3) Division of Space Mechanics and Control, National Institute of Space Research,\\
S\~ao Jos\'e dos Campos, SP 12227-010, Brazil}
\begin{abstract}
Saturn's moon Iapetus has an orbit in a transition region where the Laplace surface is bending
from the equator to the orbital plane of Saturn. The orbital inclination of Iapetus to 
the local Laplace plane is $\simeq8^\circ$, which is unexpected, because the inclination should 
be $\simeq0$ if Iapetus formed from a circumplanetary disk on the Laplace surface. It thus appears 
that some process has pumped up Iapetus's inclination while leaving its eccentricity near zero
($e\simeq0.03$ at present). Here we examined the possibility that Iapetus's inclination was 
excited during the early solar system instability when encounters between Saturn and ice giants 
occurred. We found that the dynamical effects of planetary encounters on Iapetus's orbit sensitively 
depend on the distance of the few closest encounters. In four out of ten instability cases 
studied here, the orbital perturbations were too large to be plausible. In one case, Iapetus's
orbit was practically unneffected. In the remaining five cases, the perturbations of Iapetus's 
inclination were adequate to explain its present value. In three of these cases, however, Iapetus's 
eccentricity was excited to $>$0.1-0.25, and it is not clear whether it could have been damped 
to its present value ($\simeq0.03$) by some subsequent process (e.g., tides and dynamical friction from 
captured irregular satellites do not seem to be strong enough). Our results therefore imply that 
only 2 out of 10 instability cases ($\sim$20\%) can excite Iapetus's inclination to its present 
value ($\sim$30\% of trials lead to $>$5$^\circ$) while leaving its orbital eccentricity low. 
\end{abstract}
\section{Introduction}
The moons of giant planets in the solar system can be divided into several categories.
The {\it regular} moons are large, roughly spherical satellites on nearly circular orbits
that are aligned with the host planet's equator. They are thought to have formed by
complex accretion processes in the circumplanetary disk (Canup \& Ward 2002, Mosqueira \& 
Estrada 2003). The {\it irregular} moons are smaller, irregularly shaped satellites with
large, eccentric, inclined, and often retrograde orbits. They are though to have 
been captured from heliocentric orbits (e.g., Nesvorn\'y et al. 2007). In addition,
there are the ring moons, Neptune's Triton, etc., which are not the main focus here.

Saturn's moon Iapetus has a special status among the planetary satellites. 
Its physical properties, including
the large size (diameter $D\simeq735$ km), nearly spherical figure and synchronous rotation,
are characteristic of a regular moon. Its orbit, however, is unusual in that
it is transitional between those of the regular and irregular satellites.

The regular moons, on one hand, have their orbital precession controlled by the quadrupole 
potential of the host planet's equatorial bulge, and by their mutual interaction. The 
irregular moons, instead, have their orbital precession driven by solar gravity. 
The transition between these two regimes occurs near the Laplace radius, $r_{\rm L}$, defined as:
\begin{equation}
r_{\rm L}^5=J_2'R_p^2 a_p^3(1-e_p^2)^{3/2}{M_p \over M_\odot}\ ,
\end{equation}  
where $M_p$, $R_p$, $a_p$ and $e_p$ are planet's mass, physical radius, semimajor axis and 
eccentricity, $M_\odot$ is the solar mass, and
\begin{equation}
J_2'R_p^2 = J_2R_p^2+{1 \over 2} \sum_{i=1}^n a_i^2{m_i \over M_p}\ . \label{j2prime}
\end{equation}  
Here, $J_2$ is the quadrupole coefficient, $m_i$ and $a_i$ are the mass and semimajor axis
of a satellite, and index $i$ goes over $n$ inner satellites. 

For Saturn, $J_2 \simeq 0.0163$, and for Iapetus, $J_2'=0.065$, mainly contributed by 
Titan. With the semimajor axis $a=59$ $R_p$, Iapetus is therefore just outside the Laplace 
radius, $r_{\rm L}\simeq48$ $R_p$. None of other regular or irregular 
satellites is as close to the Laplace radius. The regular satellites have $a\ll r_{\rm L}$, 
and their nodal precession is roughly uniform with respect to the planet's equator. The 
irregular satellites have $a\gg r_{\rm L}$ and their nodal precession is roughly uniform
with respect to the orbital plane of their host planet.

In the transition region, $a \sim r_{\rm L}$, the uniform precession occurs with respect to 
a surface, known as the {\it Laplace surface}, that is intermediate between the equatorial 
and orbital planes. The angle between the planetary spin axis and the normal to the Laplace 
surface, $\phi$, is defined as 
\begin{equation}
\tan 2\phi = {\sin 2 \theta \over \cos 2 \theta + 2 r_{\rm L}^5/a^5}\ ,
\label{phi}
\end{equation}
where $\theta$ is the host planet obliquity (Tremaine et al. 2009).\footnote{Note that Eqs. 
(22) and (23) in Tremaine et al. (2009) have typos (corrected, for example, in Tamayo et al. 
2013).} For Saturn's present obliquity, $\theta_{\rm S}=26.7^\circ$, and the orbital distance 
of Iapetus, this gives $\phi\simeq15^\circ$, roughly half way between the equatorial and 
orbital planes. 

\begin{figure}
\epsscale{1.0}
\plotone{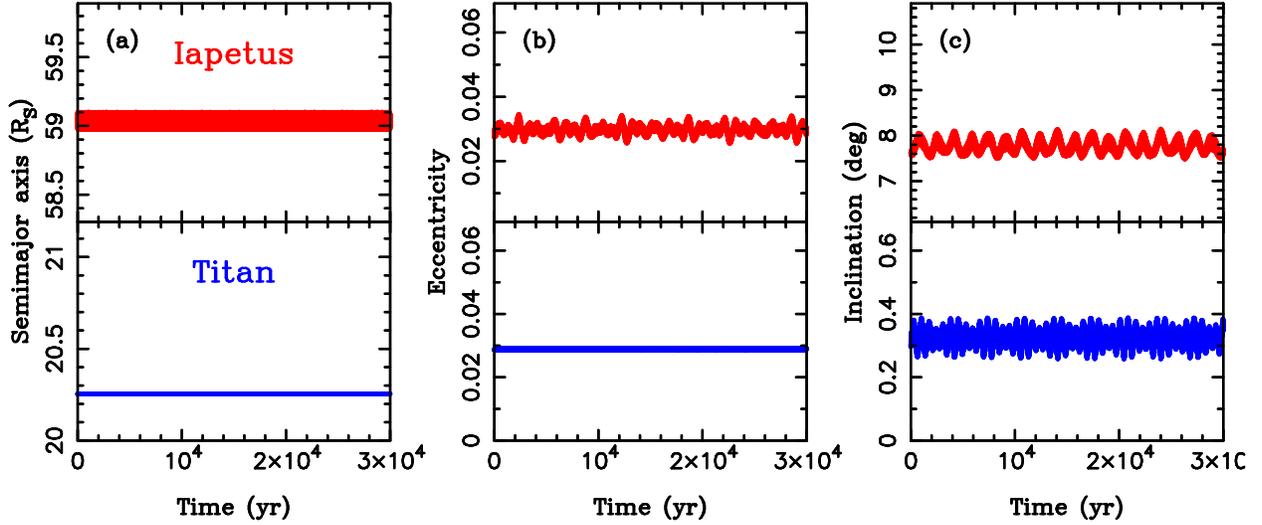}
\caption{The current orbits of Iapetus (red) and Titan (blue). From left to right the plots show
the semimajor axis, eccentricity and inclination to Laplace surface. Notably, the orbital eccentricities 
of Iapetus and Titan are similar ($\simeq0.03$). The mean orbital inclination of Iapetus to the Laplace 
surface is $\simeq8^\circ$ (panel c). To make this figure, we obtained the orbital elements of planets
and Saturn's satellites from JPL's Horizons ({\tt ssd.jpl.nasa.gov}). The orbits were numerically 
integrated from the present epoch to 30,000 yr in the future. We used a fourth-order symplectic map
(Nesvorn\'y et al. 2003). The spin axis of Saturn was kept fixed at its present value (ecliptic colatitude 
28.05$^\circ$, ecliptic longitude 79.52$^\circ$; e.g., Ward \& Hamilton 2004). The tilt of the Laplace 
surface was determined from Eq. (\ref{phi}), with a small correction for the higher gravitational moments 
and tilt of Titan's orbit with respect to Saturn's equator.}
\label{real}
\end{figure}

The mean orbital eccentricity of Iapetus is $\simeq$0.03 and its mean inclination to the 
Laplace surface is $\simeq$8$^\circ$ (Figure \ref{real}). If it formed from a circumplanetary disk, 
one would expect Iapetus to have zero eccentricity and inclination relative to this surface. It thus 
appears that some process has pumped up Iapetus's inclination while leaving its eccentricity near zero. 

Ward (1981) pointed out that the shape of the Laplace surface is affected by the mass of the 
circumplanetary disk, and suggested that the current orbit of Iapetus reflects its shape before 
the disk dispersed. However, this scenario requires a fast dispersal of the disk in $\sim10^3$~yr. 
If the dispersal were slower, the inclination relative to the Laplace surface would behave 
as an adiabatic invariant and would thus remain near zero. 

The excitation of Iapetus's inclination could have instead occurred when Saturn obtained its substantial 
obliquity. For this to work the obliquity would need to be tilted on a timescale comparable to Iapetus's 
nodal precession period (currently $P_\Omega\simeq 3440$ yr). Unfortunately, all processes proposed thus 
far to explain Saturn's obliquity act too fast (Tremaine 1991) or too slow (Hamilton \& Ward 2004, 
Ward \& Hamilton 2004) for this to be plausible.

Here we study the possibility that Iapetus's inclination was risen to its current value 
during the (hypothesized) dynamical instability 
in the outer solar system when scattering encounters of Saturn with ice giants happened (Tsiganis et al. 2005). 
Our model for dynamical perturbations of the satellite orbits in realistic instability simulations is 
explained in Section 2. The results are discussed in Section 3. In Section 4 we study the subsequent 
evolution of Iapetus's orbit to the present epoch. Section 5 concludes this paper.
\section{Model for Dynamical Perturbations of Iapetus's Orbit \\during Planetary Encounters}
Several properties of the Solar System, including the wide radial spacing and orbital 
eccentricities of the giant planets, can be explained if the early Solar System evolved through 
a dynamical instability followed by migration of planets in the planetesimal disk (Malhotra 1993, 
Thommes et al. 1999, Tsiganis et al. 2005). Recently, we developed new 
instability/migration models (Nesvorn\'y \& Morbidelli 2012; hereafter NM12), whose initial 
conditions were tightly linked to our expectations for planet formation in the protoplanetary 
nebula. We recently used these models to study the orbital behavior of the terrestrial planets
during the instability (Brasser et al. 2013), capture of Jupiter Trojans and irregular 
satellites (Nesvorn\'y et al. 2013, 2014), and survival of the Galilean satellites at Jupiter
(Deienno et al. 2014).

\begin{figure}
\epsscale{0.7}
\plotone{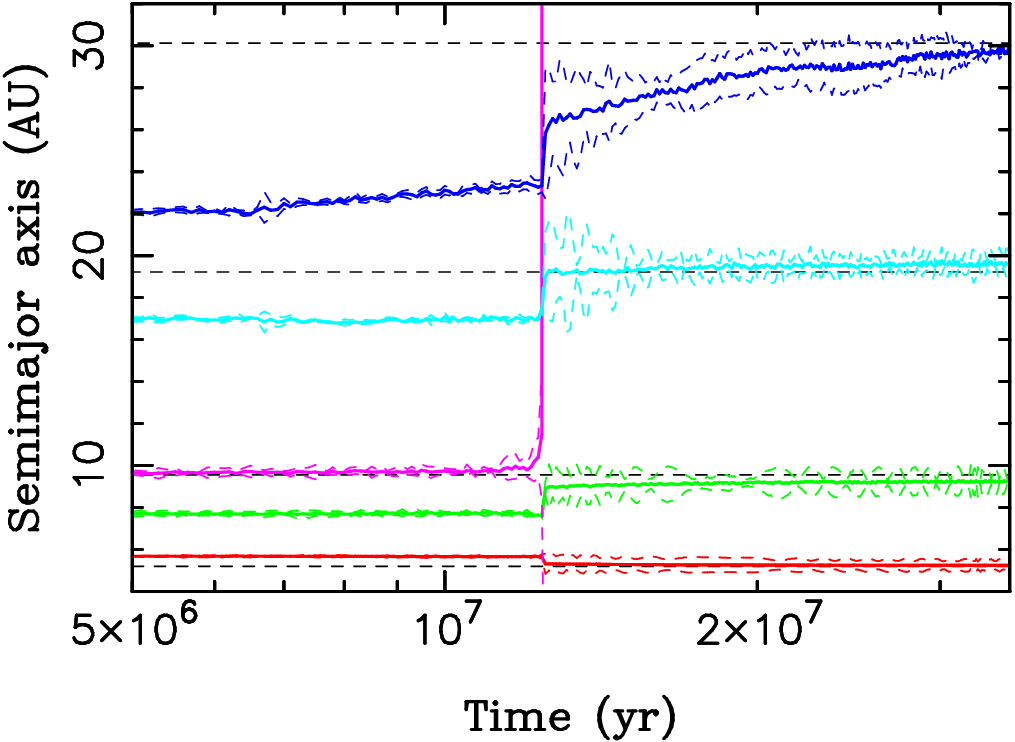}
\caption{The orbital history of the outer planets in Case 4. Jupiter, Saturn and the 3rd ice giant with mass
$M_{\rm Ice}=22\ M_\oplus$ were started in the (3:2,4:3) resonant chain. Uranus and Neptune were placed on 
non-resonant orbits with the semimajor axes 17 AU and 22.1 AU, respectively. The plot shows the 
semimajor axes (solid lines), and perihelion and aphelion distances (dashed lines) of each planet's 
orbit. The horizontal black dashed lines show the semimajor axes of planets in the present solar system. The 3rd ice
giant was ejected from the solar system at $t=1.25\times10^7$ yr after the start of the integration.}
\label{case8}
\end{figure}

Here we work with ten cases taken from NM12. Cases 1, 2 and 3 were illustrated in Nesvorn\'y et al. 
(2013, their Figures 1-4). Figure \ref{case8} shows the evolution of planets in Case 4. Table \ref{ten} 
lists the main properties of these simulations. Note that these cases were selected solely based on 
their success in reproducing the orbital properties of the solar system planets (see NM12). We therefore 
did not have any a priori knowledge of what consequences to expect in these cases for the 
satellite orbits.

In all cases considered here, the Solar System was assumed to have five giant planets initially (Jupiter, 
Saturn and three ice giants). This is because NM12 showed that having five planets initially is 
a convenient way to satisfy constraints. The third ice giant with the mass comparable to that of Uranus 
or Neptune is ejected into interstellar space during the instability (see also Nesvorn\'y 2011, Batygin 
et al. 2012). A shared property of the selected runs is that Saturn undergoes a series of encounters 
with the ejected ice giant.

For each planetary encounter (in all selected cases) we recorded the position and velocity vectors of all 
planets. Only encounters with $d<R_{{\rm H},1}+R_{{\rm H},2}$, where $d$ is the distance of planets during 
the encounter, and $R_{{\rm H},1}$ and $R_{{\rm H},2}$ are their Hill radii, were considered. The number of 
these encounters for Saturn is shown in Table \ref{ten}. In Cases 6 and 10, Saturn had the lowest (14) 
and highest (64) number of encounters, respectively. To determine the effect of these encounters on Iapetus, 
we used the recorded states and performed a second set of integrations where Iapetus and Titan were 
included. This was done as follows.

First, planets were integrated backward in time from the first encounter recorded in the previous integration.
This new integration was stopped when $d>3$ AU. At this point, Titan and Iapetus were inserted in the 
integration. We assumed that the initial orbits of both satellites were perfectly circular and on the 
Laplace surface. The semimajor axes were set to their current values ($a_{\rm T}=20.25$ $R_{\rm S}$ and 
$a_{\rm I}=59.02$ $R_{\rm S}$).\footnote{The original orbits of Titan and Iapetus before the instability are 
unknown. Ideally, we would like to start with the initial semimajor axis values that lead, after a period of 
scattering encounters, to the present ones. This is unfortunately difficult to assure in our forward modeling. 
This approximation, however, should not be a big deal in the cases where the semimajor axis changes by 
less then $\sim$10\% during scattering encounters, because the results obtained with 
slightly different initial semimajor axis values are found to be similar.} Saturn's oblateness and 
gravity of the inner moons up to Enceladus were included in 
the forward simulation via the effective quadrupole term $J_2'$ (Eq. \ref{j2prime}). The orbits of planets 
and satellites were propagated forward in time, through the encounter, and up to the point when $d>3$ AU 
again. We used the Bulirsch-Stoer integrator with a step $h=0.16$~days, which is roughly 1/100 of Titan's orbital 
period. 

Once this part was over, we removed the ice giant that participated in the encounter (to avoid any additional 
encounters during the interim period), and continued integrating the orbits of planets and satellites toward 
the next encounter. Ideally, we would like to smoothly join this integration with that corresponding
to the next encounter. This is, unfortunately, impossible with our current setup that ignores the gravitational 
effects of planetesimals (while planetesimals were included in the original simulations). Therefore, even if we 
integrated the planetary orbits all the way to the next encounter, the position and velocity vectors at that 
time would not be the same as the ones in the original simulation.

As a compromise,  we opted to respect the recorded time interval to the next encounter, $\delta t$, if
$\delta t < 3500$ yr, or integrate to $t=3500$-7000 yr if $\delta t>3500$ yr. The integration time cutoff was 
implemented to economize the CPU time. This should be correct as long as the timing of encounters is 
uncorrelated with the orbital phase of the two satellites, which should be a reasonable assumption. The 
integration was terminated randomly between $t=3500$ and 7000 yr to assure that the timing of the next 
encounter was uncorrelated with the phase of Iapetus's nodal recession.

The orbits of satellites at the end of each integration were used as the initial orbits for the next encounter.
To minimize possible discontinuities at this transition we preserved the osculating values of angles 
$\Omega_{\rm I}-\Omega_{\rm T}$ and $\Omega_{\rm I}-\Omega_{\rm S}$, where $\Omega_{\rm T}$ and $\Omega_{\rm I}$
are nodal longitudes of the two satellites with respect to the local Laplace planes, and $\Omega_{\rm S}$ is
Saturn's nodal longitude with respect to the invariant plane of the outer solar system. Angle $\Omega_{\rm I}-
\Omega_{\rm T}$ remained unchanged because we maintained the orientation of orbits of Titan and Iapetus
with respect to Saturn's orbital plane. Angle $\Omega_{\rm I}-\Omega_{\rm S}$ was preserved by applying a 
rotation on the position and velocity vectors of planets before each encounter. This procedure assured that the 
secular evolution of $i_{\rm T}$ and $i_{\rm I}$ did not suffer any artificial discontinuities during 
transitions from one encounter to another. We check on that and the discontinuities in $i_{\rm I}$ were
found to be $<0.1^\circ$, which should be insignificant.

Saturn's obliquity at the time of planetary encounters is unknown and we therefore treated it as an unknown 
parameter in our model. In each of the ten selected cases, we performed two sets of simulations with Saturn's 
obliquity $\theta_{\rm S}=0$ and $\theta_{\rm S}=26.7^\circ$. In the later case, it is assumed that Saturn's 
obliquity was excited {\it before} the instability (Ward \& Hamilton 2004, Hamilton \& Ward 2004). 
In the former case, it is assumed that Saturn's equator was aligned with the orbital plane at the 
onset of the instability. This is plausible, because it has been suggested that Saturn's current obliquity 
was excited by a secular spin-orbit resonance {\it after} the instability (Bou\'e et al. 2009). It is also 
possible that Saturn's obliquity had a value intermediate between these two extremes. We do not study these 
intermediate values here, because it turns out that satisfactory results can be obtained for both 
$\theta_{\rm S}=0$ and $\theta_{\rm S}=26.7^\circ$ (Section 4). We therefore do not have a good motivation 
to investigate the intermediate values.
 
For each case and obliquity value, we performed 1000 integrations where the initial orbital phase of satellites 
and the initial azimuthal orientation of Saturn's pole was chosen at random. These integrations should be
considered statistically equivalent, because the orbital and precessional phases at the onset of instability 
are unknown. Saturn's spin vector was assumed to be fixed in inertial space, which should be a reasonable approximation, 
because the stage of encounters typically lasts $\sim10^5$ yr, while the precession period of Saturn's spin axis
is much longer ($\simeq1.8\times10^6$ yr at present). Also, note that planetary encounters themselves cannot 
significantly change the orientation of Saturn's spin axis (Lee et al. 2007).
\section{Results of Scattering Simulations}
While the global orbital evolution of the planets was similar in the ten selected cases, the history of Saturn's 
encounters with an ice giant varied from case to case. These differences are important for perturbations of 
Iapetus's orbit and is why different cases were considered in the first place. The statistics 
of encounters is reported in Table \ref{ten}. There are between 14 (Case 6) and 64 (Case 10) recorded 
encounters in each case, a small fraction of which shows a minimal distance $d<0.1$ AU (column 5 in Table 
\ref{ten}). These very close encounters are obviously the most important for the regular satellites. The 
closest encounter of all occurred in Case 6 ($d=0.003$ AU). On the other hand, the closest encounter in 
Case 1 had $d=0.201$~AU. For reference, the semimajor axes of Titan and Iapetus are 0.0081 AU and 0.024~AU, 
respectively.  

Cases 2, 6 and 10 had at least five encounters with $d<0.1$ AU, while Case 5 had one close encounter
with $d=0.007$ AU. These cases generated very large perturbations of 
orbits of Iapetus and Titan. In most trials, Iapetus was ejected onto a heliocentric orbit. In those in which
the orbit remained bound, the eccentricity and inclination ended up implausibly large. In addition, 
perturbations of Titan's orbit often produced a very large inclination that, again, cannot be reconciled 
with the present orbit (because there is no obvious means to damp Titan's inclination back down; Section 3). 
We therefore believe that these cases are implausible. 

This is interesting because it shows that Saturn's regular satellites pose important constraints on the
instability calculations. Deienno et al. (2014) used Cases 1, 2 and 3 from NM12 discussed here and considered 
constraints from the Galilean satellites of Jupiter. They found Case 2 implausible, because perturbations 
of the orbits of the Galilean satellites were clearly excessive. This finding correlates with the 
constraints from the Saturnian satellites considered here, which also allow us to rule out Case 2.  

Table \ref{final} shows the orbital elements of Titan and Iapetus after the last planetary encounter
in Cases 1, 3, 4, 7, 8 and 9. The range of the results is broad, starting from relatively small 
perturbations such as those in Case 7, where Iapetus inclinations ended up being $2^\circ<i_{\rm I}< 3^\circ$ 
on average. Only $\simeq$10\% of trials in this case exceeded $5^\circ$. We therefore find that dynamical 
perturbations in Case 7 are not large enough to explain Iapetus's present inclination. This would imply
that Iapetus would need to have a significant inclination before the stage of encounters.

Cases 1, 3, 4, 8 and 9 appear to be more interesting. We now discuss these cases in detail. In general, 
the results show a dependence on the obliquity value of Saturn. In Case 1, for example, Iapetus's orbit reached 
inclination $i_{\rm I}=3.2_{-1.7}^{+2.0}$ deg for $\theta_{\rm S}=0$, while $i_{\rm I}=4.0_{-2.0}^{+2.6}$ deg 
for $\theta_{\rm S}=26.7^\circ$. This trend of increasing $i_{\rm I}$ with increasing $\theta_{\rm S}$
is expected because of the following arguments. With $\theta_{\rm S}=0$ the main source of the inclination 
excitation is the direct torque of the ice giant on the satellite orbit. If $\theta_{\rm S}$ is non-zero, 
however, the inclination with respect to the Laplace surface can be changed by several additional effects. 

These additional effects are {\it indirect} in that they do not change the orientation of the satellite orbital 
plane in the inertial reference frame. Instead, they affect the tilt of the Laplace surface.
A change of the semimajor axis of a satellite, for example, implies a change of $\phi$ according to Eq. 
(\ref{phi}). If the semimajor axis immediately changed back during the next planetary encounter, the original 
orbital inclination would be recovered. If, instead, the next encounter happens only after a significant 
fraction of the nodal period, satellite's orbital plane has time to recess around the Laplace surface, and the 
original inclination is not recovered. 

Both these cases occur in reality as there are typically a few dozens of encounters spread over 
$\sim10^5$ yr (while $P_\Omega =3440$ yr). Therefore, as the semimajor axis 
changes as a result of encounters, the inclination will random walk with respect to the Laplace surface. 
This effect adds to that produced on the satellite orbit by the direct torque. Additional changes
of satellite's orbital inclination relative to the Laplace surface are produced as {\it Saturn's} 
orbit is modified by scattering encounters with ice giants. They are a consequence of changes of 
Saturn's orbital inclination and the dependence of $\phi$ on Saturn's semimajor axis in Eq.~(1). 
For example, $\phi \simeq 20^\circ$ at 59 $R_{\rm S}$ for Saturn's initial semimajor axis 
$a_{\rm S}\simeq7.8$ AU (Figure \ref{case8}), while $\phi \simeq15^\circ$ with the present value
$a_{\rm S}\simeq9.55$ AU.

\begin{figure}
\epsscale{1.0}
\plotone{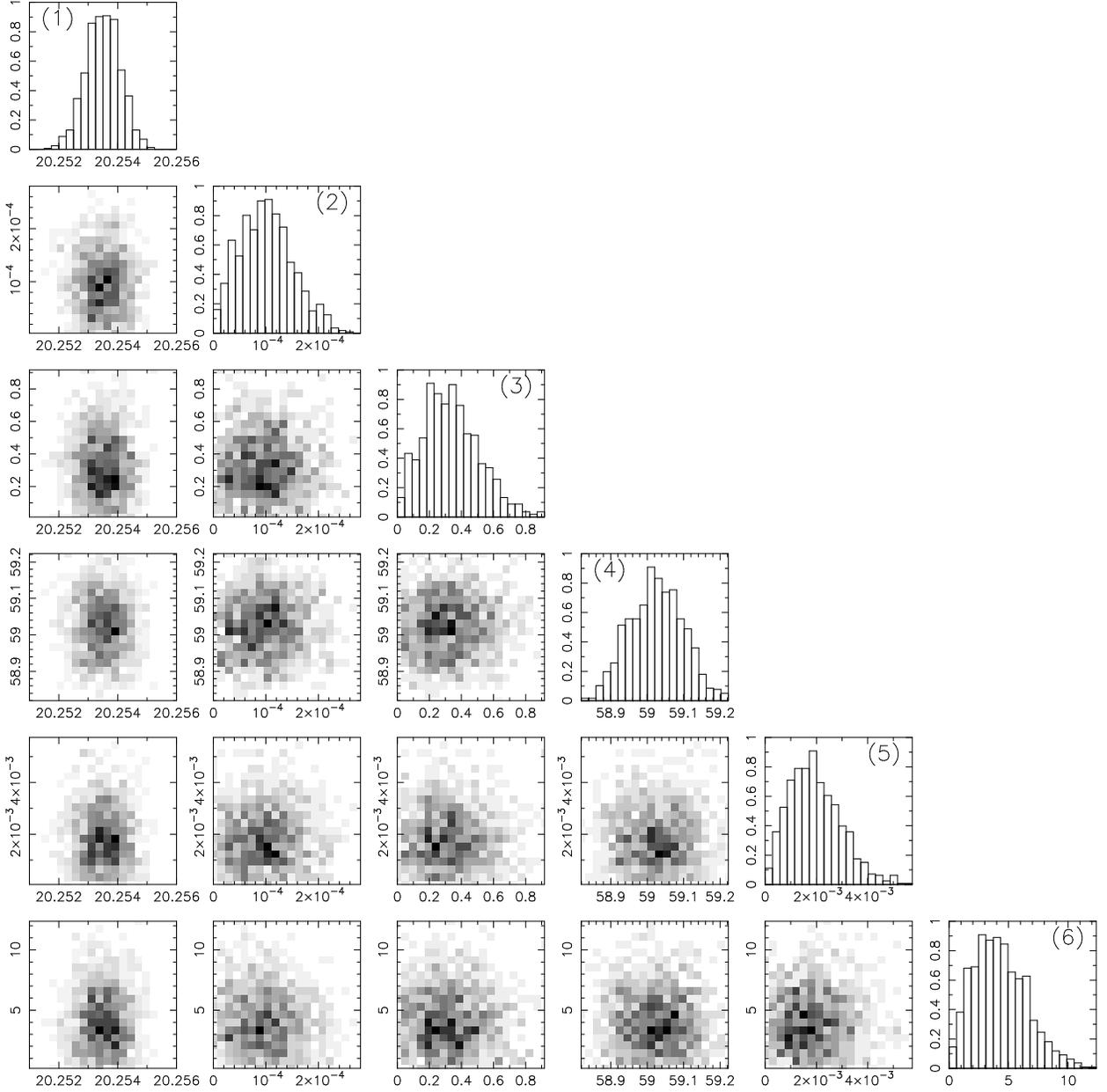}
\caption{The final distribution of orbital elements obtained in Case 1 and $\theta_{\rm S}=26.7^\circ$. The 
plots show the 1D histograms and 2D projections of: the (1) semimajor axis of Titan (the units are $R_{\rm S}$),
(2) eccentricity of Titan, (3) inclination of Titan (in degrees), (4) semimajor axis of Iapetus 
(in $R_{\rm S}$), (5) eccentricity of Iapetus, and (6) inclination of Iapetus (in degrees). 
Darker bins in the 2D projections correspond to a larger likelihood of the result. The inclinations are 
given with respect to the Laplace surface.}
\label{case1}
\end{figure}

Figure \ref{case1} shows the final distribution of orbital elements of Iapetus and Titan obtained in 
Case 1 and $\theta_{\rm S}=26.7^\circ$. This result is encouraging for several different reasons.
First, the final $i_{\rm I}>5^\circ$ in about 30\% of all trials. The probability that Iapetus 
obtained its present orbital inclination ($\simeq8^\circ$) is therefore significant. Second,  
Titan's inclination to the Laplace surface was excited to values between 0.1$^\circ$ and 0.6$^\circ$, 
with the distribution peaking at $\simeq0.4^\circ$. This is consistent with the present inclination 
of Titan ($\simeq0.34^\circ$ mean). Third, the orbital eccentricity of Iapetus remained low. There
is thus no need in this case for invoking tides or other effects to bring the eccentricity down.

\begin{figure}
\epsscale{0.9}
\plotone{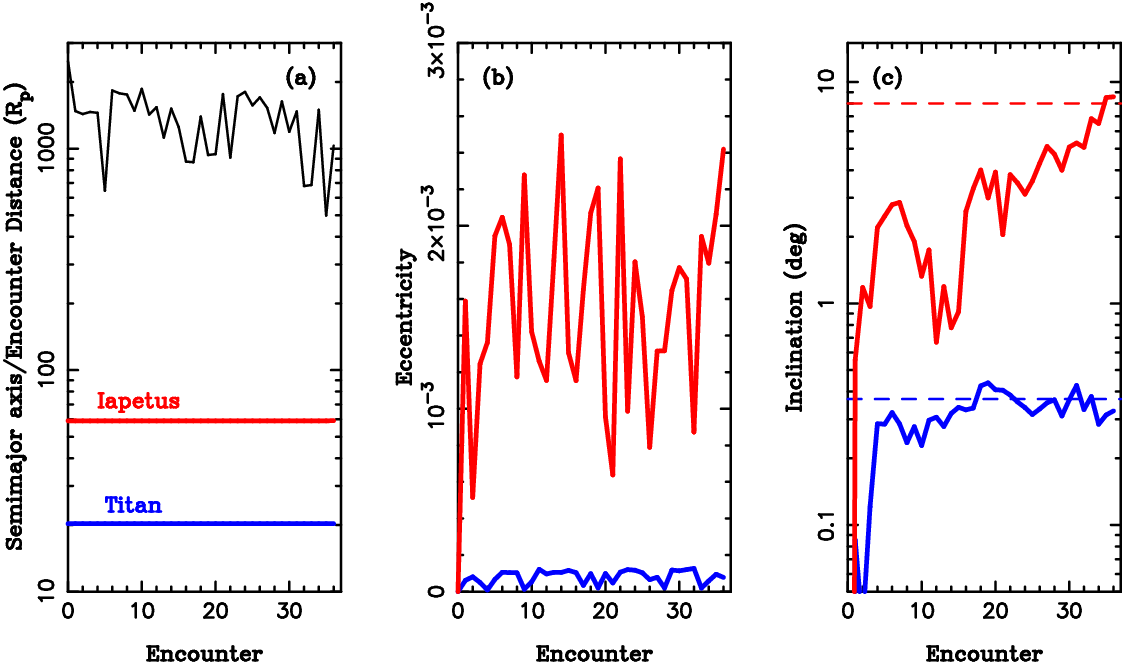}
\caption{An example of the satellite orbit evolution in Case 1. From left to right the panels show the 
semimajor axis, eccentricity and inclination of Iapetus (red) and Titan (blue). The black line in
panel (a) shows the minimal distance of Saturn and the ice giant for each encounter. The dashed 
lines in panel (c) indicate the present inclination values of orbits. Here we assumed that
$\theta_{\rm S}=26.7^\circ$. The inclinations are given with respect to the Laplace surface.}
\label{example1}
\end{figure}

Figure \ref{example1} shows the orbital elements of satellites at the end of each encounter in 
one trial integration in Case 1. This trial was selected and is shown here because it leads to 
the final values of the orbital elements that are fully consistent with the present orbits of Titan 
and Iapetus. The semimajor axis values remained nearly unchanged, eccentricities remained 
small,\footnote{This would imply that the present orbital eccentricities $\simeq$0.03 of Titan and 
Iapetus were produced by a different process that can pre-date the time of the  
planetary instability.} and inclinations reached the required values. Many trial integrations 
in Case 1 shows a similar result. 

The results in Case 1 and $\theta_{\rm S}=0$ are less ideal, because only $\simeq$15\% of trials lead to 
$i_{\rm I}>5^\circ$. This is a consequence of the general dependence of the results on Saturn's 
obliquity discussed above. We conclude that a significant initial obliquity of Saturn before the 
stage of planetary encounters helps to obtain better results in Case 1. Very similar results 
were obtained in Case 3. We therefore do not explicitly discuss Case 3 here. 

\begin{figure}
\epsscale{1.0}
\plotone{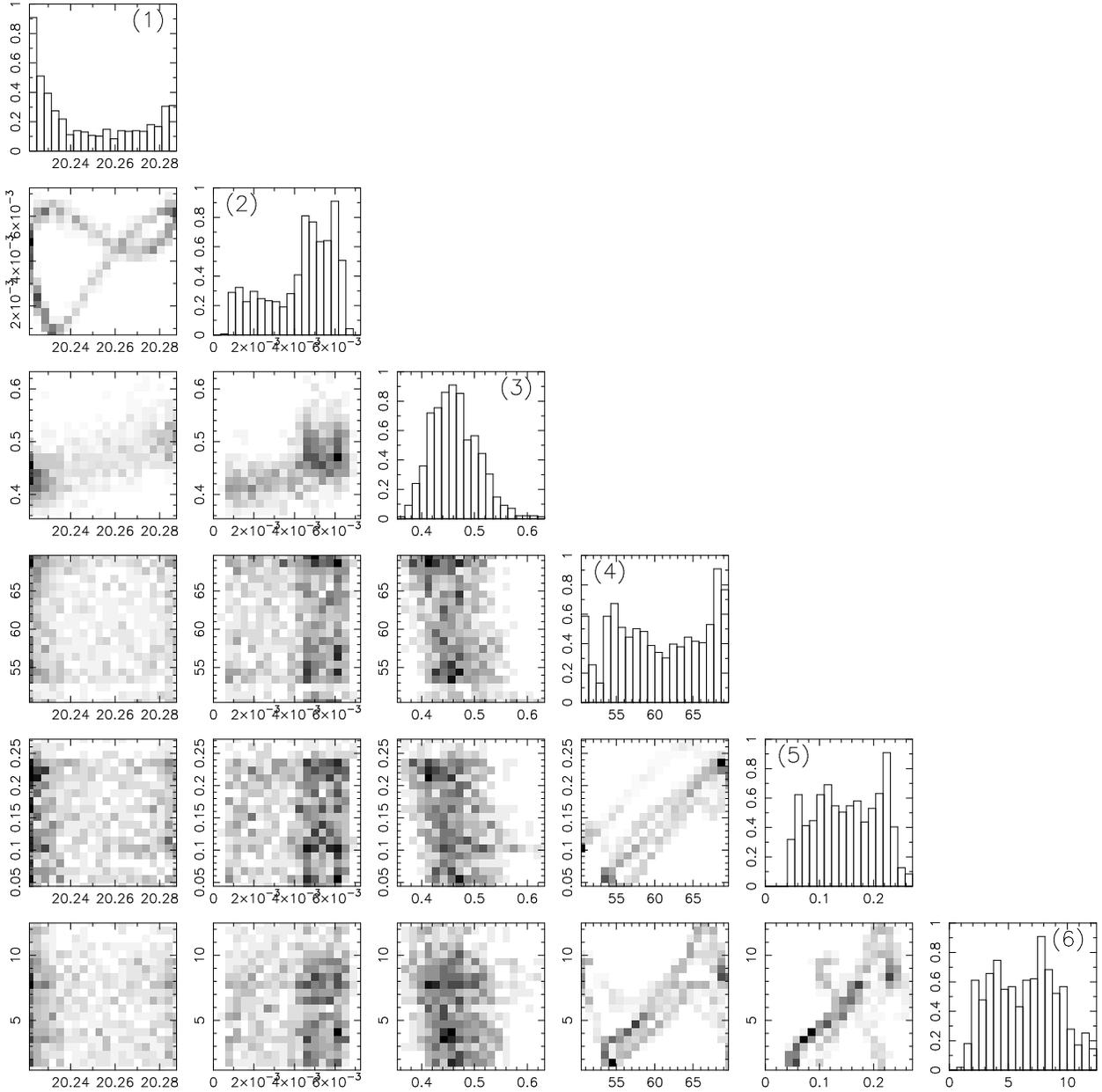}
\caption{The final distribution of orbital elements obtained in Case 4 and $\theta_{\rm S}=0$. The 
plots show the 1D and 2D projections of: the (1) semimajor axis of Titan (the units are $R_{\rm S}$),
(2) eccentricity of Titan, (3) inclination of Titan (in degrees), (4) semimajor axis of Iapetus 
(in $R_{\rm S}$), (5) eccentricity of Iapetus, and (6) inclination of Iapetus (in degrees). 
Darker bins in the 2D projections correspond to a larger likelihood of the result. The inclinations
are given with respect to the Laplace surface.}
\label{case4}
\end{figure}

Case 4 tells a different story. In this case, the closest encounter occurred at a distance 
$d=0.033$ AU, which is only $\simeq$1.3 times the semimajor axis of Iapetus. This encounter 
itself generated larger perturbation of satellite orbits than all encounters in Cases 1 and 3. 
Iapetus's inclination ended up being $6.5_{-3.3}^{+2.9}$ deg for $\theta_{\rm S}=0$ (Figure \ref{case4})
and $6.9_{-2.7}^{+3.8}$ deg for $\theta_{\rm S}=26.7^\circ$. Both these results match the present
inclination of Iapetus comfortably within 1$\sigma$. 

Interestingly, Titan's inclination was excited to $0.458_{-0.041}^{+0.048}$ deg for $\theta_{\rm S}=0$ 
and to $1.53_{-0.25}^{+0.43}$ deg for $\theta_{\rm S}=26.7^\circ$. So, based solely on this result, 
low Saturn's obliquity would be preferred in Case 4, because Titan's inclination ends up 
matching it present value better with $\theta_{\rm S}=0$ than when the obliquity is large.
This is opposite to what we have found for Cases 1 and 3 (see above).

\begin{figure}
\epsscale{0.9}
\plotone{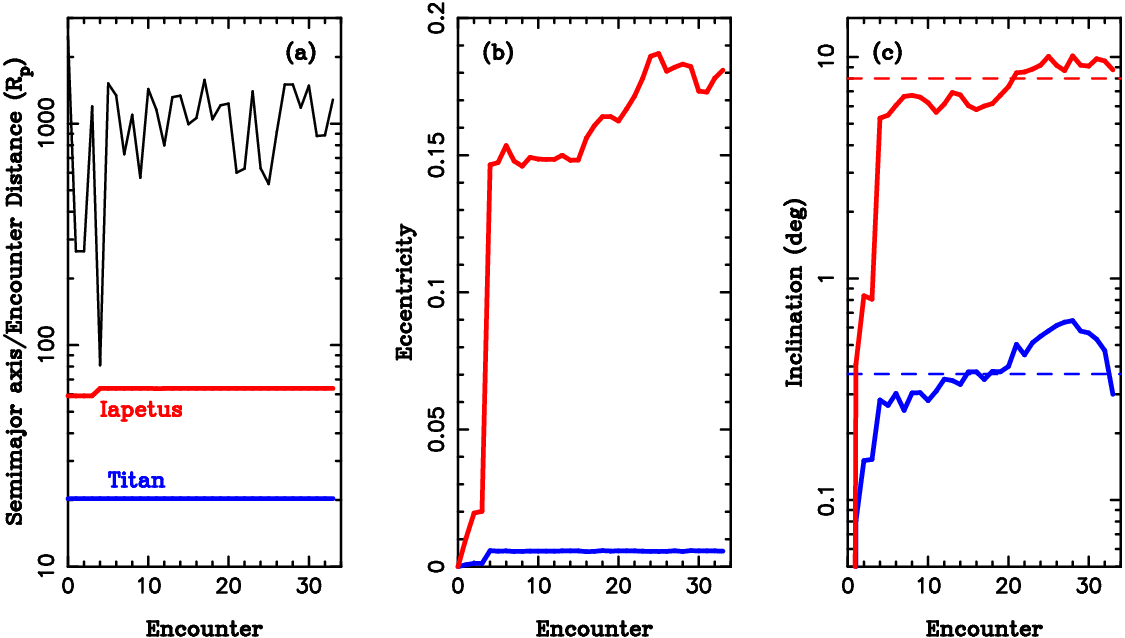}
\caption{An example of the satellite orbit evolution in Case 4. From left to right the panels show the 
semiamajor axis, eccentricity and inclination of Iapetus (red) and Titan (blue). The black line in
panel (a) shows the minimal distance of Saturn and the ice giant for each encounter. The dashed 
lines in panel (c) indicate the present inclination values of orbits. Here we assumed that
$\theta_{\rm S}=0$. The inclinations are given with respect to the Laplace surface.}
\label{example2}
\end{figure}

In addition, unlike in Cases 1 and 3, the eccentricity of Iapetus became significantly excited in 
Case 4 (to $0.151_{-0.069}^{+0.069}$ for $\theta_{\rm S}=0$ and to $0.129_{-0.059}^{+0.073}$ for 
$\theta_{\rm S}=26.7^\circ$). This case would thus require significant eccentricity damping in
times after the planetary instability (Section 4). Figure \ref{example2} illustrates the 
orbital elements of Iapetus and Titan in one trial integration in Case 4.

Cases 8 and 9 bear similarities to Case 4 (Table \ref{final}). Both these cases generated the
required excitation of Iapetus's inclination, but led to significant eccentricity that would
need to be damped after the epoch of planetary encounters. Titan's inclination obtained in these 
simulations was roughly correct, perhaps only a bit larger than what we would ideally like
(column 5 in Table \ref{final}). Titan's eccentricity remained essentially unchanged.

These results should be seen in positive light. We were only able to consider ten instability 
cases thus far. Our resolution of the initial conditions leading to the instability and planetary 
encounters is therefore grainy. Given the sensitivity of these results to the detailed history
of planetary encounters, we then find it quite possible that our tests are somewhat inadequate 
to get everything right. A more thorough investigation of parameter space will need a massive 
use of CPUs or a different approach (Vokrouhlick\'y et al., in preparation). 
\section{Subsequent Orbital Evolution of Iapetus from after \\ the Instability to the Present Epoch}
It is hypothesized that planetary instability occurred $\simeq4$ Gyr ago (e.g., Gomes et al. 2005).
Here we studied several dynamical processes that may have altered Iapetus's orbit during gigayears 
after the instability. We looked into several mechanisms: (1) dynamical effects of flybyes of 100-km class
planetesimals, (2) dynamical friction from captured irregular satellites and their debris, and (3)
tides. Our tests showed that the effects of (1) are completely negligible. Mechanism (2) would be effective
in damping Iapetus's eccentricity only if the mass captured in (or evolved to) the neighborhood of
Iapetus's orbit were $>0.1\, M_{\rm Iapetus}$, where $M_{\rm Iapetus}=1.8\times10^{21}$ kg. For comparison, the 
mass of the original population of irregular satellites captured at Saturn is estimated to be 
$\sim 10^{18}$ kg (Nesvorn\'y et al. 2014), more than two orders of magnitude lower then needed.

To study (3), we adopted a model for tides developed in Mignard (1979, 1980), where the tidal accelerations of satellites 
are given as functions of the planetocentric Cartesian coordinates  (Eqs.~(1) and (2) in Lainey et~al. 2012). 
The tidal evolution of satellite's orbit was studied by direct numerical integrations of orbits with a symplectic 
$N$-body code known as {\tt swift\_rmvs3} (Levison \& Duncan 1994) that we modified to include Mignard's 
tidal acceleration terms. The satellite rotation was assumed to be synchronous.\footnote{To implement the 
synchronous rotation in the code we adopted the following approximation (V. Lainey, personal communication). 
For tides raised on a satellite, we only included the radial component of the acceleration. This component does not 
depend on satellite's rotation rate, and is therefore independent of the detailed assumptions about 
synchronicity. We then multiplied the magnitude of this component by 7/3 to account for the effects of 
the longitudinal component of the tidal acceleration. This is because in the limit of small eccentricities, 
which is applicable here, the orbital energy dissipated by radial flexing of the satellite is $3/4$ of that 
dissipated by satellite's librations (e.g., Murray \& Dermott 1999, Chapter~4).}
The dissipation effects were parametrized by the standard tidal parameter $Q' = Q/k_2$, where $k_2$ is the 
quadrupole Love number and $Q$ is the quality factor (assumed constant here). As for Saturn, previous 
theoretical work indicated the $Q'$ values at least of the order of $10^4$  (e.g., Peale et al. 1980, Zhang
\& Nimmo 2009), but Lainey et al. (2012) recently suggested from the astrometric modeling of Cassini's 
observations that $Q' \simeq 4350$ (with about 30\% uncertainty). We use this later value but point out that 
our main conclusions (related to the eccentricity of Iapetus) are essentially independent of Saturn's $Q'$. 

Instead, the strength of tidal damping of the eccentricity of Iapetus sensitively depends, via the 
secular orbital coupling of Iapetus to other moons,\footnote{Iapetus itself is too far from Saturn 
for direct tides to be important for Iapetus's orbital evolution.} on the dissipation of tidal energy 
in Titan and the inner satellites. We find that it is problematic to quantify this process, because 
$Q'$ values of Saturn's satellites are poorly known. We therefore performed several numerical integrations 
with widely ranging values of $Q'$. All satellites between Mimas and Iapetus were included. 

\begin{figure}
\epsscale{1.0}
\plotone{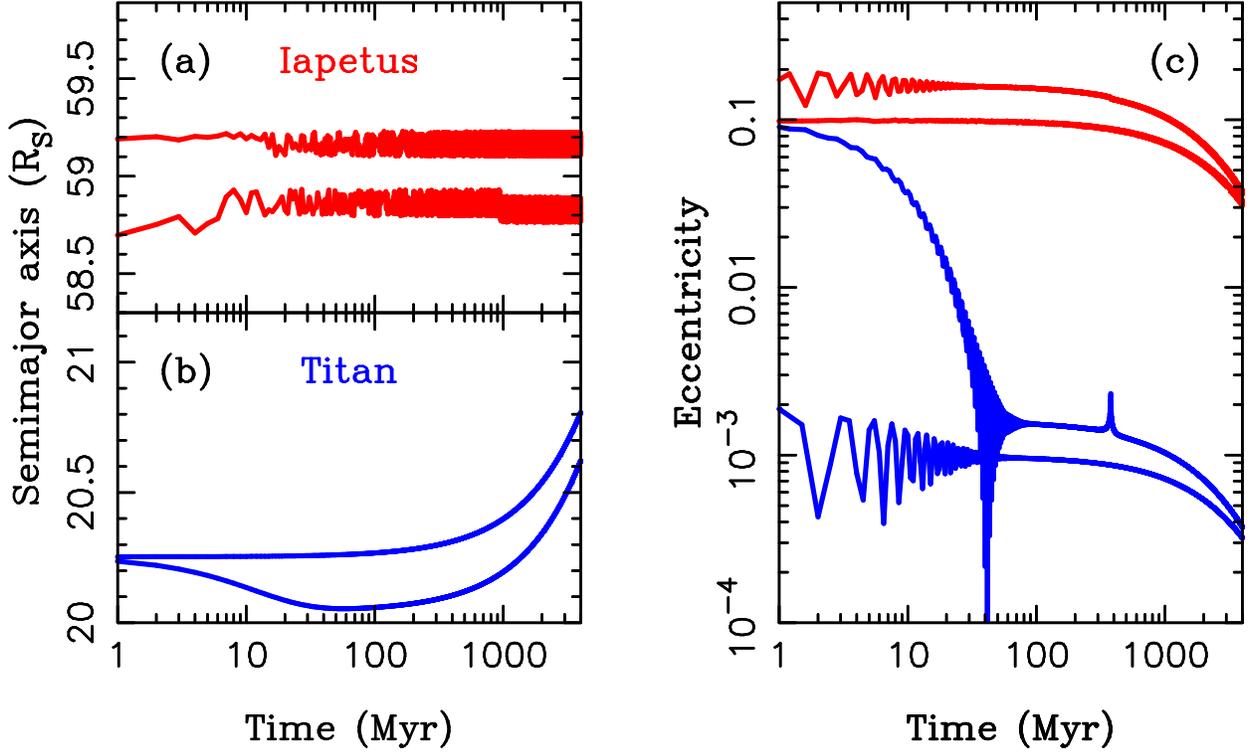}
\caption{Examples of the orbital evolution of Titan (blue) and Iapetus (red) due to tides. 
Two cases with different initial setups are shown here: (1) $e_{\rm T}=0$ and $e_{\rm I}=0.1$, 
and (2) $e_{\rm T}=0.1$ and $e_{\rm I}=0.2$. In panel (a), the two tracks of Iapetus were 
offset by $\pm0.2\ R_{\rm S}$ for clarity. The orbital inclinations of Titan and Iapetus, not 
shown here, do not practically change over 4~Gyr. The final eccentricity of Iapetus, after 
4 Gyr of tidal evolution, is roughly equal to the present value ($e_{\rm I} \simeq0.029$). 
In both cases, we fixed $Q'_{\rm S}=4350$ (Lainey et al. 2012) and used the same value 
of $Q'$ for all satellites, $Q'=20$ and $Q'=10$, respectively, for the two initial setups. 
The extremly strong tidal dissipation in satellites implied by these values is probably 
implausible (see the main text). Note that Titan and Iapetus were started with their present 
value of the semimajor axis in these simulations, which is strictly speaking incorrect.
See \'Cuk et al. (2013) for a more detailed analysis of the orbital history of Titan and 
Iapetus.} 
\label{tides}
\end{figure}

The principal result of these integrations is that the secular coupling of Iapetus to Titan and the 
inner moons, and the tidal dissipation in Titan and the inner moons, would potentially be capable 
of reducing Iapetus's eccentricity only if $Q' \lesssim 20$ (Figure \ref{tides}). This low value of $Q'$, 
however, appears to be implausible. For example, dynamical constraints suggest that $Q'\simeq10^3$-$10^4$ 
for Enceladus and Dione (Zhang \& Nimmo 2009). For Titan, theory indicates that the strength of tidal 
dissipation should sensitively depend on its interior structure (e.g., Sohl et~al. 1995), but is unlikely 
as low as needed here. Measurements elsewhere in the solar system suggest $Q'\simeq70$ for Io 
(Lainey et al. 2009) and $Q'\simeq1200$ for the Moon (Khan et~al. 2004). 

In summary, we find that if the orbital eccentricity of Iapetus would have become excited during planetary 
encounters, it would probably stay high to the present epoch. This result has important implications for 
the interpretation of our scattering experiments (Section 3), because it shows that only 3 out of 10 instability
cases considered here (numbers 1, 3 and 7 in Table 2) appear to be plausible based on Iapetus's 
eccentricity constraint. We also find that Titan's and Iapetus's orbital inclinations, if excited by planetary 
encounters, would remain essentially unchanged during the subsequent tidal evolution. Titan's inclination 
(current mean $i_{\rm T}=0.34^\circ$) does not appear to be a problem because it stays low during the 
scattering phase in all three cases mentioned above. While Case 7 would require that Iapetus's 
inclination was excited already before the scattering phase, Cases 1 and 3 are capable of generating
Iapetus's inclination during the scaterring phase (Figure \ref{example1}). 
\section{Conclusions}
The orbital inclination of Iapetus is a long standing problem in planetary science. The inclination 
should be $\simeq0$ if Iapetus formed from a circumplanetary disk on the Laplace surface, but it 
presently is $\simeq8^\circ$. Here we investigated the possibility that Iapetus attained its significant 
orbital inclination during a hypothesized instability in the outer solar system when Saturn had close
encounters with an ice giant. We found that roughly 50\% of instability cases that satisfy other 
constraints (see NM12) are capable of exciting Iapetus's (and Titans's) inclination to the present 
value. For most of these cases to be plausible, however, some dissipation mechanism is required to 
damp the orbital eccentricity of Iapetus that is typically excited by encounters to $>$0.1. 
In only 2 out of 10 instability cases studied here, the eccentricity of Iapetus remained low while
the orbital inclination of Iapetus was significantly excited (such that $i_{\rm I}>5^\circ$ in at 
least 30\% of trials; Cases 1 and 3). These different outcomes depend on the number and minimum
distance of the encounters, and on their geometry.

One of our main motivations for this study was the question of whether it is possible to have a history
of encounters between Saturn and an ice giant that leads to capture of the irregular satellites
at Saturn via the mechanism described in Nesvorn\'y et al. (2007) {\it and} to satisfy constraints from 
Saturn's regular satellites. Here we demonstrated that it is indeed possible to satisfy these 
constraints simultaneously (e.g., in Cases 1 and 3; see Nesvorn\'y et al. (2014) for irregular satellite 
capture in these cases). Moreover, we found that the orbital perturbation of the regular satellites 
mainly results from a few closest encounters. The results are therefore expected to be highly variable. 
As a rough criterion, we find that the closest encounter of the ice giant to Saturn cannot be closer 
than 0.05 AU or about 2 times the semimajor axis of Iapetus (or 0.02 AU if the eccentricity constraint 
is relaxed). Capture of irregular satellites, on the other hand, mainly depends on the bulk of distant 
encounters, and is expected to occur generically. The regular and irregular satellites thus represent 
somewhat different, and not mutually exclusive constraints. 

Our final remark is related to the orbital perturbations of regular satellites at Jupiter, Uranus and 
Neptune. Deienno et al. (2014) already demonstrated that the orbits of the Galilean satellites remain 
unchanged in Case 3 studied here, while they suffer implusibly large excitations in Case 2. Also, 
according to Deienno et al., from the perspective of the Galilean moons, the Case 1 is intermediate 
between Cases 2 and 3. Our tests for Jupiter, using the same methodology as described for Saturn
in Section 2, confirm these results and show, in addition, that Cases 4, 7, 8, and 10 generate 
only modest (and plausible) perturbations of the Galilean satellite orbits. Therefore, while there
is a hint of correlation of the results for Jupiter and Saturn, there are also cases such as 
the Case 10, where the Galilean moons survive essentially undisturbed while Saturn's regular 
satellites, including Titan, plunge in disorder. These cases could give the right framework for the 
hypothesis of the late origin of the Saturn system (Asphaug \& Reufer 2013).

Interestingly, the satellites of Uranus are a very sensitive probe for planetary encounters. This is because
the most distant of these satellites, Oberon, has the semimajor axis comparable to Rhea in the Saturnian
system, and only $\simeq$0.068$^\circ$ inclination with respect to the Laplace surface. Previous works
done in the framework of the original Nice model and the jumping-Jupiter model with four planets 
(Deienno et al. 2011; Nogueira et al. 2013; and R. Gomes, personal communication) had difficulties to 
satisfy this constraint, because Uranus experienced many encounters with Jupiter and/or Saturn
in these instability models. In the models taken from NM12, however, Uranus does not have encounters 
with Jupiter and Saturn, and instead experiences a relatively small number of encounters with 
a relatively low-mass ice giant. According to our tests, Oberon's inclination remains below 0.1$^\circ$ in all 
cases studied here, except for Case 8. This result could be used to favor the NM12 instability 
models. Neptune's satellites are less of a constraint in this context, because Triton's orbit is closely 
bound to Neptune and has been strongly affected by tides (Correia et al. 2009).

\acknowledgments
This work was supported by NASA's Outer Planet Research program. The work of D.V. was partly supported 
by the Czech Grant Agency (grant P209-13-013085). R. D. was supported by FAPESP (grants 2010/11109
and 2012/23732). We thank Matija \'Cuk for a very helpful referee report.

\clearpage
\begin{table}[t]
\begin{center}
\begin{tabular}{lrrrrr}
\hline
\# & $M_{\rm Ice}$ & $M_{\rm disk}$ & \# of enc. & \# of enc. & $d_{\rm min}$ \\
   & ($M_\oplus$)  & ($M_\oplus$)  &            & in 0.1AU  &   (AU)     \\ 
\hline
{\bf 1} &  15    &  20            &   36       &    0       & 0.201       \\
2       &  15    &  20            &   51       &    5       & 0.021       \\
{\bf 3} &  15    &  20            &   29       &    2       & 0.056       \\
{\bf 4} &  22    &  20            &   33       &    2       & 0.033       \\
5       &  22    &  22            &   22       &    1       & 0.007       \\
6       &  22    &  22            &   14       &    5       & 0.003       \\
{\bf 7} &  18    &  20            &   27       &    1       & 0.057       \\
{\bf 8} &  18    &  20            &   17       &    1       & 0.020       \\
{\bf 9} &  18    &  20            &   15       &    3       & 0.054       \\
10      &  18    &  22            &   64       &    6       & 0.038       \\
\hline
\end{tabular}
\end{center}
\caption{The basic properties of ten selected instability cases from NM12. The columns 
are: (1) id number, (2) mass of the 3rd ice giant, (3) mass of the planetesimal disk,
(4) number of encounters between Saturn and ice giants, (5) number of encounters
with $d<0.1$~AU, and (6) distance of the closest encounter. Plausible cases that 
do not lead to excessive perturbations of the orbits of Iapetus and Titan are denoted in 
bold in column 1. In Cases 1, 2 and 3, the 
five planets were placed in a resonant chain (3:2,3:2,2:1,5:3). In Cases 4 to 10,
Jupiter, Saturn and the 3rd ice giant were placed in a resonant chain (3:2,4:3), and
Uranus and Neptune on non-resonant orbits with semimajor axes 17 AU
and 22.1 AU, respectively. The planetesimal disk with mass indicated in column 3
was represented by 1000 massive disk particles initially located between 24 and 30 
AU. See NM12 for additional information on these runs.}
\label{ten}
\end{table}

\clearpage
\begin{table}[t]
\begin{center}
\begin{tabular}{lrllllll}
\hline 
\# & $\theta$ ($^\circ$) & $a_{\rm T}$ ($R_{\rm S}$) & $e_{\rm T}$ & $i_{\rm T}$ ($^\circ$) & $a_{\rm I}$ ($R_{\rm S}$) & $e_{\rm I}$ & $i_{\rm I}$ ($^\circ$) \\
   &                    & 20.25                    &  0.029      & 0.34                 &   59.02                   & 0.029      &                 8.1  \\     
1  & 0 & $20.2535_{-0.0007}^{+0.0007}$    & $0.00010_{-0.00005}^{+0.00007}$ & $0.17_{-0.09}^{+0.11}$ 
       & $59.03_{-0.09}^{+0.07}$          & $0.002_{-0.001}^{+0.001}$ & $3.2_{-1.7}^{+2.0}$ \\
1  & 26.7 & $20.2535_{-0.0006}^{+0.0006}$ & $0.00009_{-0.00005}^{+0.00005}$ & $0.33_{-0.16}^{+0.17}$ 
          & $59.02_{-0.07}^{+0.08}$      & $0.002_{-0.001}^{+0.001}$       & $4.0_{-2.0}^{+2.6}$ \\     
3  & 0    & $20.2534_{-0.0003}^{+0.0004}$ & $0.0012_{-0.0002}^{+0.0002}$ & $0.15_{-0.07}^{+0.08}$ 
          & $58.3_{-1.9}^{+1.8}$            & $0.094_{-0.044}^{+0.071}$       & $2.4_{-1.2}^{+1.3}$ \\     
3  & 26.7 & $20.2535_{-0.0004}^{+0.0003}$ & $0.0006_{-0.0002}^{+0.0003}$ & $0.46_{-0.13}^{+0.14}$ 
          &  $58.4_{-2.1}^{+1.7}$           & $0.095_{-0.045}^{+0.075}$       &  $3.6_{-1.8}^{+2.0}$ \\  
4  & 0    & $20.24_{-0.01}^{+0.04}$      & $0.005_{-0.003}^{+0.001}$     & $0.46_{-0.04}^{+0.05}$ 
          & $60.6_{-6.4}^{+7.4}$            & $0.151_{-0.069}^{+0.069}$       & $6.5_{-3.3}^{+2.9}$ \\  
4  & 26.7 & $20.25_{-0.02}^{+0.02}$       & $0.0004_{-0.002}^{+0.004}$    & $1.5_{-0.3}^{+0.4}$
          & $60.3_{-6.5}^{+7.2}$            & $0.129_{-0.059}^{+0.073}$       & $6.9_{-2.7}^{+3.8}$ \\      
7  & 0    & $20.2533_{-0.0005}^{+0.0006}$    & $0.0005_{-0.0002}^{+0.0002}$  & $0.16_{-0.07}^{+0.008}$
          & $59.01_{-0.24}^{+0.25}$          & $0.008_{-0.004}^{+0.009}$    & $2.0_{-0.9}^{+1.1}$ \\  
7  & 26.7 & $20.2533_{-0.0009}^{+0.0009}$    & $0.0004_{-0.0002}^{+0.0003}$     & $0.30_{-0.15}^{+0.18}$ 
          & $59.04_{-0.26}^{+0.23}$         & $0.008_{-0.005}^{+0.007}$     & $2.8_{-1.4}^{+1.6}$ \\ 
8  & 0    & $20.33_{-0.08}^{+0.04}$           & $0.030_{-0.012}^{+0.012}$        & $0.74_{-0.12}^{+0.13}$
          & $61_{-11}^{+21}$            & $0.233_{-0.047}^{+0.085}$           & $6.6_{-2.1}^{+3.4}$ \\
8  & 26.7 & $20.26_{-0.08}^{+0.10}$      & $0.030_{-0.017}^{+0.025}$        & $0.91_{-0.70}^{+0.59}$ 
          & $63_{-13}^{+20}$            & $0.248_{-0.066}^{+0.087}$       & $9.0_{-5.1}^{+5.1}$ \\   
9  & 0    & $20.2532_{-0.0005}^{+0.0005}$    & $0.0037_{-0.0006}^{+0.0009}$     & $1.15_{-0.06}^{+0.05}$ 
          & $55.5_{-2.9}^{+2.9}$            & $0.24_{-0.11}^{+0.14}$        & $5.9_{-2.1}^{+3.6}$ \\  
9  & 26.7 & $20.2532_{-0.0004}^{+0.0004}$    & $0.004_{-0.001}^{+0.002}$      & $1.9_{-0.8}^{+0.2}$
          & $55.8_{-2.5}^{+3.0}$            & $0.24_{-0.10}^{+0.12}$         & $8.0_{-4.1}^{+4.3}$ \\     
\hline
\end{tabular}
\end{center}
\caption{The orbital elements of Titan and Iapetus obtained in our scattering simulations. The columns
show: (1) id number, (2) Saturn's obliquity ($\theta_{\rm S}$), (3)-(5) Titan's semimajor axis,
eccentricity and inclination ($a_{\rm T}$, $e_{\rm T}$ and $i_{\rm T}$), (6)-(8) Iapetus's semimajor axis, 
eccentricity and inclination ($a_{\rm I}$, $e_{\rm I}$ and $i_{\rm I}$). The inclinations are given 
relative to the Laplace surface. The values listed here are the medians and their $\pm$34.1\% associated 
quantiles. The first row with numbers lists the current mean orbital elements of Titan and Iapetus.}
\label{final}
\end{table}

\clearpage

\clearpage

\clearpage

\clearpage

\clearpage

\clearpage

\clearpage

\end{document}